\documentclass[EPiC]{easychair}

\usepackage{doc}
\usepackage{amsmath}
\usepackage{amssymb}
\usepackage[backend=biber,style=ieee, sorting=none]{biblatex}
\bibliography{easychair}

\title{A Hybrid Physical–Digital Framework for Annotated Fracture Reduction Data Evaluated using Clinically Relevant 3D metrics
}

\author{
   Basile Longo\inst{1,2}
\and
   Paul-Emmanuel Edeline\inst{1,3}
\and
    Hoel Letissier\inst{1,2,4}
\and   
   Marc-Olivier Gauci\inst{5, 6}
\and   
   Aziliz Guezou-Philippe\inst{1,3}
\and
   Valérie Burdin\inst{1,3}
\and 
   Guillaume Dardenne\inst{1}
}

\institute{
  LaTIM INSERM U1101,
  Brest, France
\and
    Université de Bretagne Occidentale, France
\and
   IMT Atlantique, Brest, France
\and
   CHU de Brest, France
\and 
    Institut Universitaire Locomoteur et du Sport, Hôpital Pasteur II, CHU de Nice, France
\and
    Equipe ICARE, INSERM U1091, Université Côte d'Azur, France\\
  \email{basile.longo@univ-brest.fr}
 }

\authorrunning{Longo et al.}

\titlerunning{A Hybrid Physical–Digital Framework for Annotated Fracture Reduction Data}

\begin{document}

\maketitle

\begin{abstract}
A major bottleneck in Computer-Assisted Preoperative Planning (CAPP) for fracture reduction is the limited availability of annotated data. While  annotated datasets are now available for evaluating bone fracture segmentation algorithms, there is a notable lack of annotated data for the evaluation of automatic fracture reduction methods. Obtaining precise annotations, which are essential for training and evaluating automatic CAPP algorithm, of the reduced bone therefore remains a critical and underexplored challenge. Existing approaches to assess reduction methods rely either on synthetic fracture simulation which often lacks realism, or on manual virtual reductions, which are complex, time-consuming, operator-dependant and error-prone. 

To address these limitations, we propose a hybrid physical-digital framework for generating annotated fracture reduction data. Based on fracture CTs, fragments are first 3D printed, physically reduced, fixed and CT scanned to accurately recover transformation matrix applied to each fragment. 

To quantitatively assess reduction quality, we introduce a reproducible formulation of clinically relevant 3D fracture metrics, including 3D gap, 3D step-off, and total gap area. The framework was evaluated on 11 clinical acetabular fracture cases reduced by two independent operators. Compared to preoperative measurements, the proposed approach achieved mean improvements of $168.85 ~\mathrm{mm^2}$  in total gap area, $1.82 ~\mathrm{mm}$ in 3D gap, and $0.81 ~\mathrm{mm}$ in 3D step-off.

This hybrid physical–digital framework enables the efficient generation of realistic, clinically relevant annotated fracture reduction data that can be used for the development and evaluation of automatic fracture reduction algorithms.

\end{abstract}

\section{Introduction}
\label{sect:introduction}

Computer-Assisted Preoperative Planning (CAPP) for fracture reduction and fixation has gained popularity due to its effectiveness in reducing blood loss, operative time, and intraoperative fluoroscopic use \cite{moolenaar2022computer}. Its primary goal is to improve the surgeon’s understanding of fracture patterns in order to achieve improved osteosynthesis outcomes \cite{mensel2022preoperative, moolenaar2022computer}. Advances in computer science and deep learning have further transformed CAPP from 2D radiographic planning into more advanced 3D visualization and analysis tools \cite{mensel2022preoperative, moolenaar2022computer}.

Despite these advances, modern CAPP approaches remain highly data-intensive. A major bottleneck in trauma research is the limited availability of annotated data \cite{yibulayimu2025fracformer}. While considerable efforts have been devoted to fractured bone segmentation—particularly for complex cases such as pelvic fractures \cite{PENGWIN}—little attention has been paid to the acquisition of annotated data for fracture reduction. Such data are essential for the training and evaluation of automatic bone reduction methods. Existing approaches either rely on synthetically generated fractures derived from healthy bones \cite{zeng2024bidirectional}, which often produce unrealistic fracture patterns, or on manual virtual reductions \cite{yibulayimu2025fracformer} which are extremely time-consuming, highly operator-dependent, and prone to ambiguities arising from 2D–3D visualization.

To address these limitations, we propose a novel method for acquiring annotated fracture reduction data. Bone fragments are 3D printed, physically reduced using a hot glue gun, and then CT scanned. This process allows surgeons to interact with and reduce fractures in a more intuitive and engaging manner. To quantitatively evaluate the resulting reductions, we also propose 3D measurements adapted from Meesters et al. \cite{meesters2019introduction}.


\section{Material and Methods}
\label{sect:Material_and_Methods}

\subsection{Data}

Using the publicly available PENGWIN \cite{PENGWIN} and CLINIC6 (a fracture-specific subset of CTPelvic1k \cite{CTPelvic1k}) datasets, both originally designed for segmentation tasks, we selected only acetabulum fractures.

A total of 11 acetabulum fracture cases were selected for this study, 5 from CLINIC6 and 6 from PENGWIN, with a mean number of 3.5 fragments (min-max: 2-6), all with an intact contralateral acetabulum. According to the Letournel acetabulum fracture system \cite{letournel1993surgical}, the fractures were classified as follows: 2 Anterior wall, 1 Posterior column, 2 Transverse, 3 Both columns, 2 T-shaped, and 1 Transverse with posterior wall.

\subsection{3D Printing and Manual Reduction}

Each bone fragment was 3D printed in polylactic acid (PLA) by the hospital’s 3D-printing unit (W.Print, Brest CHU, France). Following preliminary testing, a hot glue gun was chosen as the adhesive for fragment assembly, as it provides sufficient mechanical stability for handling, sets rapidly, and allows fragments to be detached and repositioned if necessary.

Manual reductions were performed twice, by medical experts, after which the assembled models were CT scanned to obtain fragment-wise reference transformation matrices.

\subsection{Segmentation quality}
Because the models consist solely of plastic components without surrounding soft tissues, the individual reduced fragments can be quickly and reliably semi-automatically segmented using 3D Slicer\footnote{https://www.slicer.org/}. The accuracy of the semi-automatic segmentation method is evaluated by ensuring that the post-reduction reconstruction is close to the original one. This is quantified for all fragments using the Chamfer distance ($d_{\mathrm{CD}}$) between the original reconstructed mesh $M_o$ and the physically reduced reconstructed mesh $M_r$, defined as
\[
d_{\mathrm{CD}}(M_o, M_r)
=
\frac{1}{|M_o|}
\sum_{\mathbf{x} \in M_o}
\min_{\mathbf{y} \in M_r} \|\mathbf{x} - \mathbf{y}\|
+
\frac{1}{|M_r|}
\sum_{\mathbf{y} \in M_r}
\min_{\mathbf{x} \in M_o} \|\mathbf{y} - \mathbf{x}\|,
\]
where $|M_o|$ and $|M_r|$ are the cardinal of each points cloud (mesh) and $\|\cdot\|$ the Euclidean distance. 


\begin{figure}[tb]
    \label{fig:fig1}
    \centering
        \centering
        \footnotesize
        \includegraphics[width=1.0\linewidth]{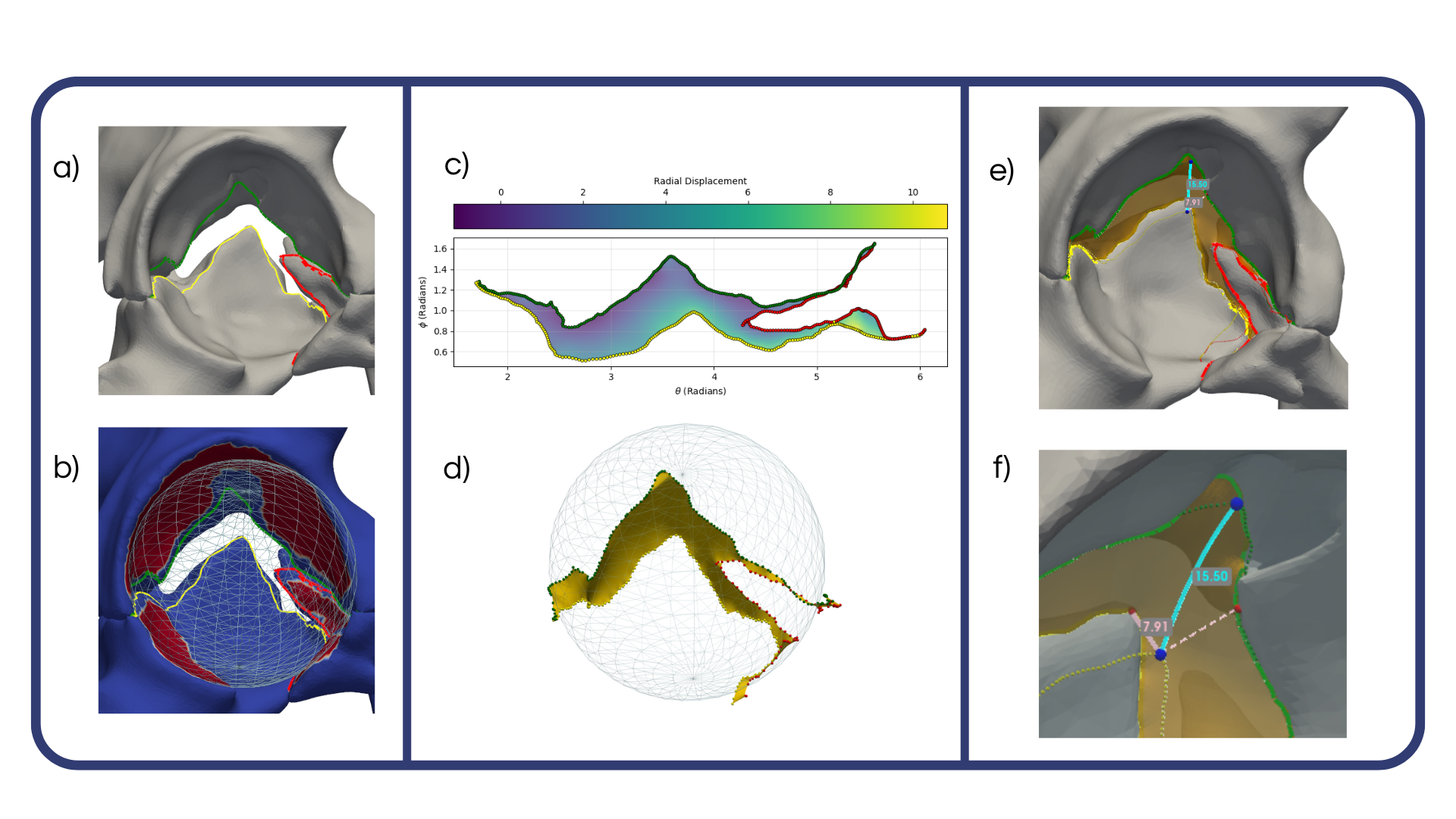}

    \caption{Total gap area generation. (a) Fracture lines identification. (b) Articular surface labeling and sphere fitting (c) Fracture lines projection onto the sphere, 2D surface generation and per point radial interpolation. (d) Reprojection to 3D space. (e) Total gap area generated in orange. (f) 3D gap in cyan and 3D step-off in pink; continuous lines are the original fracture lines and dotted lines are their projection.}
\end{figure}

\subsection{Reduction quality}

To evaluate reduction quality, we adapt the 3D CT-based measurements proposed by Meesters et al. \cite{meesters2019introduction}, which extends classical 2D "gap and step-off" metrics into a 3D context. This also includes the total gap area, i.e. the surface between fracture lines, which serves as a critical predictor for hip survivorship and the transition to total hip arthroplasty \cite{meesters2022quantitative}.

To semi-automatically compute these scores, we approximate the acetabular articular surface as a sphere \cite{allaire2007robust,cereatti2010human}, fitted to manually selected points on the lunate surface. Manually labeled fracture lines are then projected onto this geometry, allowing for precise 3D measurements:
\begin{itemize}
    \item 3D Gap: The maximum geodesic distance between projected fracture lines.
    \item 3D Step-off: The maximum radial deviation from the spherical projection.
\end{itemize}   

To compute the total gap area, the projected lines are unwrapped into a 2D angular domain. Within this domain, the internal surface is reconstructed and the radial deviation is interpolated using radial basis functions (RBF). Finally, the surface is mapped back to 3D to quantify the total defect, as illustrated in \hyperref[fig:fig1]{Figure 1}.

\section{Results}

\paragraph{Segmentation quality}

A mean reconstruction error of $1.36 \pm 0.24~\mathrm{mm}$ has been obtained between the original and reduced meshes.



\paragraph{Reduction quality}
Reductions were evaluated using the previously described 3D metrics: 3D gap area, 3D gap, and 3D step-off. Results are summarized in Table~1.
Operator~1 obtained a mean 3D gap area of $519.43 \pm 390.33~\mathrm{mm^2}$, a 3D gap of $8.03 \pm 3.94 ~\mathrm{mm}$, and a 3D step-off of $6.55 \pm 2.67~\mathrm{mm}$. 
Operator~2 obtained $460.90 \pm 347.33~\mathrm{mm^2}$, $6.68 \pm 3.26~\mathrm{mm}$, and $5.93 \pm 2.87~\mathrm{mm}$, respectively.
Overall, the operator achieved a mean improvement of $168.85 \pm 316.11~\mathrm{mm^2}$ in total gap area, $1.82 \pm 3.95 ~\mathrm{mm}$ in 3D gap, and $0.81 \pm 3.75 ~\mathrm{mm}$ in 3D step-off compared with preoperative measurements.

\begin{table}[h!]
\centering
\renewcommand{\arraystretch}{1.3}
\begin{tabular}{|c|c|c|c|}
\hline
                & Total gap area (mm$^2$) & 3D gap (mm)  & 3D step-off (mm) \\
\hline
Preoperative    & 659.02 ± 356.42        & 9.18 ± 4.10  & 7.05 ± 3.80      \\
\hline
Operator 1      & 519.43 ± 390.33        & 8.03 ± 3.94  & 6.55 ± 2.67      \\
\hline 
Operator 2      & 460.90 ± 347.33        & 6.68 ± 3.26  & 5.93 ± 2.87      \\
\hline
\end{tabular}
\caption{3D measurements for the preoperative situation and after reduction for both operators.}
\label{tab:example}
\end{table}

\paragraph{Inter-Observer Agreement}
Inter-observer variability between the two operators was evaluated for the 3D metrics. 
For the 3D gap area, the mean difference was $58.53 \pm 151.02~\mathrm{mm^2}$ 
For the 3D gap, the mean difference was $1.35 \pm 1.82~\mathrm{mm}$ 
For the 3D step-off, the mean difference was $0.62 \pm 1.92~\mathrm{mm}$ 

\section{Discussion}

We proposed an intuitive and fast method to accurately collect annotated data of fracture reductions, along with quantitative metrics to evaluate reduction quality. 

A bidirectional Chamfer error $1.36 \pm 0.24~\mathrm{mm}$ indicates that the segmentation is close to the original one.

The reduction done by the experts delivered a mean improvement 168.85 mm² in total gap area, 1.82 mm in 3D gap, and 0.81 mm in 3D step-off.

In Meesters et al., 3D gap and step-off measurements rely on the displacement of corresponding fracture lines relative to a reference reduction, either preoperatively or postoperatively. In contrast, our inter–fracture-line measurements do not require any reference reduction, enabling a direct assessment of the fracture state. Additionally, Meesters et al. compute the total gap area directly using CAD software, whereas we model the hip joint as a sphere—a common approximation \cite{allaire2007robust, cereatti2010human}—to obtain a more reproducible approach. Although alternative geometries such as conchoids or ellipsoids may better represent hip anatomy \cite{5627704, cerveri2014patient}, these could be explored in future work to further refine the measurements.

In addition, other anatomical bone joints can be described as a mathematical surfaces and this metric approach can be transposed for other fracture, with for instance distal radius or femur articular surfaces associated to an ellipsoid \cite{allaire2007ellipsoid}. 

A limitation of this data collection method is that, in cases of minimally displaced fractures, the 3D printing approach may be less accurate than the original anatomical configuration. This reduced accuracy is primarily attributable to the additional thickness introduced by the adhesive layer between fragments, as well as the technical challenges associated with assembling and handling multiple fragments in highly comminuted fractures, other mounting techniques could be explored. Alternative mounting techniques could be investigated to address these issues.

\section{Acknowledgments}
This work was supported by the French government via the National Research Agency (ANR),
under the references ANR-23-CPJ1-0131-01 and ANR-23-RHUS-0011. The authors gratefully thank Samuel Guigo (W.Print, Brest CHU) and the PLaTIMed platform (LaTIM, Brest, France) for providing 3D-printed models and CT imaging data, respectively.

\label{sect:bib}
\printbibliography

\end{document}